\documentclass[anonymous, conference]{IEEEtran}
\usepackage{cite}
\usepackage{amsmath,amssymb,amsfonts}
\usepackage{algorithmic}
\usepackage{graphicx}
\usepackage{textcomp}
\usepackage{xcolor}
\usepackage{multirow}
\usepackage{tabularx}
\usepackage{soul}
\usepackage{xcolor}
\usepackage{graphicx}
\usepackage{subcaption}
\usepackage{times} 
\usepackage{caption} 
\usepackage[hidelinks]{hyperref}
\usepackage[numbers, square]{natbib}

\makeatletter
\def\NAT@def@citea{\def\@citea{\NAT@separator}}
\makeatother

\definecolor{lightyellow}{rgb}{1,1,0.7}
\definecolor{lightgreen}{rgb}{0.7,1,0.7}

\sethlcolor{lightyellow}  %
\def\BibTeX{{\rm B\kern-.05em{\sc i\kern-.025em b}\kern-.08em
    T\kern-.1667em\lower.7ex\hbox{E}\kern-.125emX}}
\begin{document}

\title{\fontsize{14pt}{16pt}\selectfont \textbf{Combining Retrieval and Classification: Balancing Efficiency and Accuracy in Duplicate Bug Report Detection}}

\author{\IEEEauthorblockN{1\textsuperscript{st} Qianru Meng}
\IEEEauthorblockA{\textit{LIACS} \\
\textit{Leiden University}\\
Leiden, Netherlands \\
mengqr@vuw.leidenuniv.nl}
\and
\IEEEauthorblockN{2\textsuperscript{nd} Xiao Zhang}
\IEEEauthorblockA{\textit{CLCG} \\
\textit{University of Groningen}\\
Groningen, Netherlands \\
xiao.zhang@rug.nl}
\and
\IEEEauthorblockN{3\textsuperscript{rd} Guus Ramackers}
\IEEEauthorblockA{\textit{LIACS} \\
\textit{Leiden University}\\
Leiden, Netherlands \\
g.j.ramackers@liacs.leidenuniv.nl}
\and
\IEEEauthorblockN{4\textsuperscript{th} Visser Joost}
\IEEEauthorblockA{\textit{LIACS} \\
\textit{Leiden University}\\
Leiden, Netherlands \\
j.m.w.visser@liacs.leidenuniv.nl}
}

\maketitle

\begin{abstract}
In the realm of Duplicate Bug Report Detection (DBRD), conventional methods primarily focus on statically analyzing bug databases, often disregarding the running time of the model. In this context, complex models, despite their high accuracy potential, can be time-consuming, while more efficient models may compromise on accuracy. To address this issue, we propose a transformer-based system designed to strike a balance between time efficiency and accuracy performance. The existing methods primarily address it as either a retrieval or classification task. However, our hybrid approach leverages the strengths of both models. By utilizing the retrieval model, we can perform initial sorting to reduce the candidate set, while the classification model allows for more precise and accurate classification. In our assessment of commonly used models for retrieval and classification tasks, sentence BERT and RoBERTa outperform other baseline models in retrieval and classification, respectively. To provide a comprehensive evaluation of performance and efficiency, we conduct rigorous experimentation on five public datasets. The results reveal that our system maintains accuracy comparable to a classification model, significantly outperforming it in time efficiency and only slightly behind a retrieval model in time, thereby achieving an effective trade-off between accuracy and efficiency. 

\textbf{\textit{Keywords-Duplicate Bug Detection; Deep Learning; Natural Language Processing; Transformer; Running Time; Accuracy.}}

\end{abstract}

\section{Introduction\label{sec:intro}}
Bug reports are crucial in the software development and maintenance phase, providing valuable information to software developers \cite{gupta2021systematic}\cite{Neysiani2020irvsml}. It commonly comprises structured text (e.g., timestamp, version, component, and bug status) and unstructured text, such as title and description \cite{Messaoud2022Bert}. Typically, bugs are recorded in the Bug Database (also known as Bug Tracking System) by developers, testers and users \cite{Messaoud2022Bert}\cite{zhang2016lr}\cite{rocha2021siamese}. Unfortunately, the inconsistent understanding of bug descriptions by different writers leads to the continuous generation of numerous duplicate bug reports \cite{kukkar2020duplicate}, which increases maintenance costs. Consequently, significant research efforts have been devoted to detecting duplicate bugs, aiming to reduce redundant work involving the testing of bugs that have already been resolved \cite{rocha2021siamese}\cite{runeson2007detection}\cite{Uddin2017bugprioritization}, thereby enhancing the efficiency of the bug fixing process \cite{nguyen2012duplicate}.

\begin{table}[h]
\centering
\caption{COMPARISON OF DUPLICATE BUG REPORTS FROM ECLIPSE}
\begin{tabular}{|p{1.5cm}|p{6cm}|}
\hline
Bug\_id & 178 \\
\hline
Title & \sethlcolor{lightyellow}\hl{Maintain sync view expansion state} when \sethlcolor{lightyellow}\hl{switching modes}\\
\hline
Description & It would be nice if when things got filtered out, their \sethlcolor{lightgreen}\hl{expansion} would be \sethlcolor{lightgreen}\hl{remembered}, so that when the item is revealed again it has the correct \sethlcolor{lightgreen}\hl{expansion}. For example, if you have one outgoing change; \sethlcolor{lightgreen}\hl{switch} to the catchup pane and then come back, the tree is completely collapsed. \\
\hline
Bug\_id & 226 \\
\hline
Title & \sethlcolor{lightyellow}\hl{Switching} between sync UI modes should \sethlcolor{lightyellow}\hl{preserve expansion state} \\
\hline
Description & When you \sethlcolor{lightgreen}\hl{switch} between Catch Up and Release modes, it loses the \sethlcolor{lightgreen}\hl{expansion state} of the tree. It should \sethlcolor{lightgreen}\hl{remember} this and probably the selection and top item (scroll bar position) as well. \\ 
\hline
\end{tabular}
\label{comparison_table}
\end{table}

DBRD task can be defined as: the automatic process of identifying and comparing the semantic content in bug reports to discover new reports that are duplicate or highly similar to existing reports. As shown in Table \ref{comparison_table}, there are two instances of duplicate bug reports where similar features have been highlighted. These features are not limited to exact word matching, but also extend to semantic similarity and context. Therefore, this places high demands on the capacity of automatic text processing techniques.

Traditionally, the automatic approach to DBRD has been divided into two distinct tasks: Information Retrieval (IR) and classification\cite{gupta2021systematic}. Early methods for IR primarily relied on word-based approaches (e.g., Vector Space Model), as well as topic-based models like Latent Dirichlet Allocation (LDA) and Latent Semantic Analysis (LSA), which transformed bug reports into feature vectors. More recently, embedding models, such as Word2Vec \cite{xie2018detecting}\cite{he2020duplicate}, GloVe \cite{pankajakshan2021glove}, and sentence BERT \cite{isotani2021duplicate} have gained traction. These models generate embeddings that are then utilized to calculate similarities between bugs, typically using distance measurements, such as Cosine similarity. These retrieval methods have demonstrated promising performance, particularly in terms of recall rate \cite{gupta2021systematic}. Simultaneously, classification models, particularly deep-learning-based approaches, have emerged as prominent research focus in DBRD \cite{gupta2021systematic}\cite{rocha2021siamese}\cite{nguyen2012duplicate}. Initially, the classifier employed the Convolutional Neural Network (CNN) \cite{kukkar2020duplicate}\cite{he2020duplicate}\cite{deshmukh2017towards}\cite{xiao2020hindbrrnn}, followed by the Recurrent Neural Network (RNN) \cite{xiao2020hindbrrnn} and eventually transitioning to the Long Short-Term Memory (LSTM) model \cite{xiao2020hindbrrnn}. However, due to the challenges associated with processing lengthy text, the performance of these three models has been surpassed in recent years by transformer-based classifiers, most notably BERT \cite{Messaoud2022Bert}\cite{rocha2021siamese}, sentence BERT \cite{Messaoud2022Bert}\cite{isotani2021duplicate} and RoBERTa \cite{Messaoud2022Bert}. These large language models are pre-trained on large corpus and fine-tuned on domain-specific data, which enables them to capture contextual semantic information and generate word and sentence representations efficiently. Not surprisingly, transformer-based models became the state-of-the-art for this task \cite{Messaoud2022Bert}.

However, in previous studies, we found that commonly used dataset splitting methods have data leakage issues, which may lead to biased results. Specifically, it is possible for a single data within a pair in the train set to be combined with another data and consequently appear in the development or test set. This unintentional leakage has not been explicitly addressed by most existing methods, with only one work taking this matter into account without explicitly acknowledging it \cite{Messaoud2022Bert}. Therefore, one of the main contributions of our work is the design of a Cluster-based dataset partition mechanism to address this problem.

Most importantly, while there has been a considerable emphasis on performance metrics, such as recall and precision in existing studies, the evaluation of these approaches' efficiency in terms of speed has often been overlooked. As highlighted by Haruna et al. \cite{isotani2021duplicate} in their research, with the advent of large language models, such as BERT, the performance of retrieval and classification tasks has shown remarkable advancements. Nevertheless, the deployment and execution of these models can present difficulties due to their relatively slower inference times. Especially when it comes to practical applications, the speed plays a critical role. As a result, it's essential to evaluate a model not just based on its accuracy, but also on its efficiency.

In our research, we propose a novel system based on the transformer architecture that combines the advantages of retrieval model and classification model. Our approach integrates retrieval techniques to retrieve an initial set of potential duplicate instances, which is then fed into a classification model for further triage. This innovative methodology enables us to achieve faster performance without compromising accuracy. By effectively merging these two components, we attain a balance between efficiency and accuracy in DBRD task.\\

The contributions of our work are as follows:

\begin{itemize} 
\item Cluster-based dataset partition mechanism: To address the problem of train set leakage, we introduce a cluster-based dataset partition mechanism. This mechanism ensures that duplicate instances are evenly distributed across the train and test sets, effectively mitigating any potential data leakage issues.
\item Comparison with previous models: We conduct a comprehensive comparison between the performance of the transformer-based models and that of previous methods in retrieval and classification. Through rigorous experiments and evaluations, we demonstrate that our transformer-based models outperform on both tasks, surpassing the performance of previous models.
\item Integration of retrieval model and classification: Our proposed system leverages the strengths of both retrieval models and classification models. As demonstrated in the experiments, our system can achieve a balanced between speed and accuracy in two real-world scenarios.
\end{itemize}

We introduce the related work in Section \ref{sec:background}, detail our approach in Section \ref{sec:methodology} and validate the experiments in large open source projects to demonstrate the effectiveness in Section \ref{sec:experiment} and Section \ref{sec:result}.

\section{Related work\label{sec:background}}



As previously mentioned, solutions to DBRD can be viewed as IR task and classification task. Approaches to IR tasks focus on identifying duplicates by computing similarities between textual representations, while classification tasks typically utilize deep learning techniques to train models in distinguishing between "duplicated" and "non-duplicated" instances based on learned patterns. In the following subsections, we present related work on these methods.

\subsubsection{Information Retrieval Methods}
Hiew \cite{hiew2006assisted} introduces a retrieval method for unstructured text including titles and descriptions. Textual fields are converted to TF-IDF vectors, which are then organized into clusters based on their similar characteristics to identify duplicates. Runeson et al. \cite{runeson2007detection} utilized a Vector Space Model to present text-based information and determined the text similarity by using three similarity calculation methods. Wang et al. \cite{wang2008ir} integrated execution data into their strategy to detect similar bug reports. Sun et al. \cite{sun2011towards} proposed a REP model that incorporates similarity of lexical features and categorical features from bug reports. Nguyen et al. \cite{nguyen2012duplicate} introduced the DBTM model that processes topic features extracted by LDA model and unstructured textual features. It combines topic model and retrieval model to show both similarity and dissimilarity between bug reports. Some follow-up studies \cite{alipour2013contextual}\cite{lazar2014improving}\cite{aggarwal2017detecting} adopt a similar approach to previous studies, also implementing topic models for retrieval, but differentiate their studies by analyzing distinct corpora and utilizing varied feature inputs in bug reports.

Therefore, traditional IR methods primarily focus on the calculation of word frequency feature to detect duplicates, which show advantages in processing structured text and keyword-based queries. However, IR methods exhibit limitations in processing contextual information and complex semantic features, areas where deep learning (DL) methods demonstrate proficiency.

\subsubsection{Deep Learning Methods}
Deshmukh et al. \cite{deshmukh2017towards} were the first to introduce deep learning into duplicate bug report detection, proposing a model that uses Siamese Convolutional Neural Networks and Long Short Term Memory to process hybrid input from bug reports for retrieval and classification. Budhiraja et al. \cite{budhiraja2018dwen}\cite{budhiraja2018towards} proposed Deep Word Embedding Networks (DWEN), a framework designed to retrieve similar reports by processing unstructured input, including bug report titles and descriptions. Xie et al. \cite{xie2018detecting} introduced a deep learning framework named DBR-CNN, which enhances traditional CNN by integrating domain-specific features extracted from bug reports. The hybrid features are fed into the CNN model to obtain concatenated vectors, which are utilized for classification task. Poddar et al. \cite{poddar2019train} proposed a neural architecture for multi-task learning, with joint tasks of classifying duplicates and clustering latent topics, operating on unstructured descriptions as input. Building upon the CNN framework, He et al. \cite{he2020duplicate} subsequently developed a Dual-Channel CNN (DC-CNN) method to classify duplicate bug reports using hybid-structured text as input. Kukkar et al. \cite{kukkar2020duplicate} presented a deep learning based classification model applied on hybrid features, also leveraging CNN to extract relevant features that are subsequently used to compute similarities for classification purposes. 

Following these advancements, transformer-based language models have gained considerable attention and popularity within the present landscape of duplicate bug report detection, due to their rich context-based learning capabilities. Isotani et al. \cite{isotani2021duplicate} introduced transformer-based deep learning embedding model of SBERT to vectorize the unstructured textual features (title and description) and then computes the similarity of the embedding representations, enabling retrieval of similarly ranked bug reports. Rocha et al. \cite{rocha2021siamese} proposed a SiameseQAT approach, using BERT and MLP to concatenate structured and unstructured features and features extracted based on corpus topics for retrieval and classification tasks respectively. Messaoud et al. \cite{Messaoud2022Bert} proposed a BERT-MLP model for classifying duplicate bug reports, which considers only unstructured data. The model utilizes BERT to generate contextualized word representations and applies an MLP for classification. Jiang Y et al. \cite{jiang2023does} suggested a CombineIRDL method, which utilizes different deep learning models to extract lexical, categorical, and semantic features from hybrid input and then employs a retrieval model to obtain ranked duplicates.

Building on these deep learning methods discussed in the literature, we find that most of them accomplish duplicate detection by implementing retrieval and classification tasks separately \cite{rocha2021siamese}\cite{kukkar2020duplicate}\cite{deshmukh2017towards} or focus on a single task \cite{Messaoud2022Bert}\cite{xie2018detecting}\cite{he2020duplicate}\cite{isotani2021duplicate}\cite{budhiraja2018dwen}\cite{budhiraja2018towards}\cite{poddar2019train}\cite{jiang2023does}. Furthermore, deep learning methods have demonstrated significant effectiveness in both tasks. In IR tasks, deep learning enhances similarity assessments by employing advanced word embedding models, such as transformer models. In classification tasks, it trains models (such as CNN, LSTM or transformer models) to predict whether two bug reports are duplicates by leveraging their learning capabilities to discern complex textual patterns. By employing these advanced deep learning models, classification tasks can achieve higher accuracy than IR tasks through the extraction of comprehensive textual features, but at the cost of thousands of computations to achieve such precision. Conversely, IR tasks can achieve more significant efficiency in reducing the search space than classification tasks. However, previous work has not considered the trade-off between accuracy and efficiency. Therefore, our approach combines those two tasks in order to fully exploit their strengths in terms of efficiency and accuracy, thereby achieving a balance. In doing so, we apply transformer-based models in our approach, which are widely recognized as the state-of-the-art for Natural Language Processing (NLP) tasks by exploiting their ability to learn semantic and contextual information. These models are utilized to generate embedding representations in the retrieval task and to identify duplicate pairs in the classification task. The following section contains more details of our methodology.

\section{Methodology\label{sec:methodology}}
\begin{figure*}[h]
  \centering  \includegraphics[width=\linewidth]{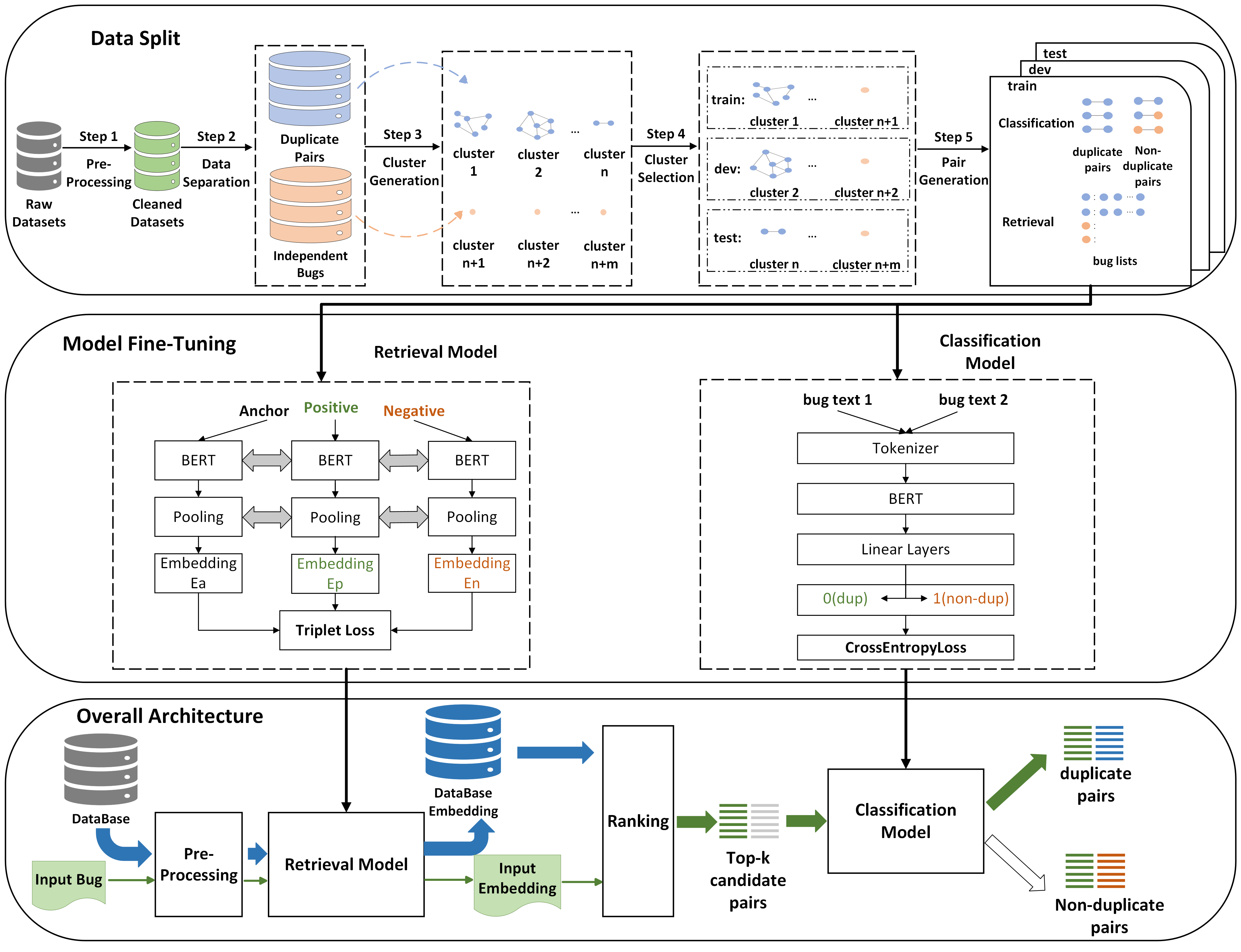}
  \caption{Transformer-Based Framework split into Two Phases: 1). Data Generation Phase and 2). Model Fine-Tuning Phase}
  \label{Pipeline}
\end{figure*}
{In this section, we outline our methodology for the DBRD task, including the overall architecture, pre-processing,  data split, and model fine-tuning.}
\subsection{Overall Architecture}
The overall architecture of our proposed approach is shown in Figure \ref{Pipeline}, which consists of three important phases: data split, retrieval and classification.
\begin{itemize}
    \item {\textbf{Data split} is to split the test and train set for preparing for the training of retrieval and classification models and we introduce it detailed in Section \ref{subsec: datasplit}}.
    \item {In the \textbf{Retrieval} phase, the retrieval model is responsible for generating the embedding representation of bug report input. Using cosine similarity \cite{rahutomo2012cosine}, it calculates the embedding similarity and selects the Top K similar bug reports. This "ranking" is primarily used to identify the top-K candidates, which serve as the target input for the subsequent classification model.}
    \item {In the \textbf{Classification} phase, the model takes the top-K candidates from the retrieval phase and aims to output the final results by labeling them as duplicates or non-duplicates.}
\end{itemize}


\subsection{Pre-processing}
In DBRD task, the input of bug reports can typically be unstructured input containing only unstructured textual fields, or hybrid input including both unstructured textual features and structured categorical features. Our emphasis lies on unstructured input, specifically title and description. These text constitute the most critical component of bug reports and they are also noisy and complex, covering a large number of domain-specific technical fields. 
To eliminate the redundant and invalid data in the content of bug report datasets, the following operations are used to clean the datasets:
\begin{itemize}
    \item Remove all non-English words from the text (note: adjust this based on the primary language of the dataset)
    \item Remove some special characters but keep periods and commas
    \item Remove stop words
    \item Unify all letters to lowercase
\end{itemize}

\subsection{Data Split} \label{subsec: datasplit}

As outlined in Section \ref{sec:intro}, most prior studies have neglected the leakage from the train set to the development and test set. To address this issue, we introduce a cluster-based dataset generation method, as illustrated in the Data Split Phase of Figure \ref{Pipeline}. This approach ensures a stringent separation between the train set, development set, and test set.

\textbf{Data Separation.} After pre-processing, the bug reports are categorized into two groups: independent bug reports and duplicate bug reports. Independent bug reports refer to those that do not have any duplicates among the dataset.

\textbf{Cluster Generation.} In this particular step, we adopt the assumption of transitivity in the relationship between duplicate bug reports. This means that if Bug A is a duplicate of Bug B, and Bug B is a duplicate of Bug C, then Bug A and Bug C are also considered duplicates. Consequently, based on our example, Bugs A, B, and C would be grouped together in a single distinct cluster. Leveraging this assumption, we employ a cluster-based method to group pairs of duplicated bug reports into distinct clusters. Within each cluster, every two bugs are marked as duplicates. 

\textbf{Cluster Selection.} Unlike prior studies that randomly select duplicate pairs for train/test sets, we choose clusters to form train, development, and test sets. This ensures each bug appears only once in any set, and the clusters across these sets are distinct. This method reduces potential biases and data leakage from directly selecting duplicate pairs.


\textbf{Pair Generation.} In this step, we generate two train/test datasets with different data structures for retrieval and classification models respectively. The retrieval dataset follows the format of $[$Bug ID: Bug IDs$]$, where bugs sharing the same cluster are considered duplicates of each other. On the other hand, the classification dataset comprises both duplicated pairs and non-duplicated pairs in a one-to-one format of $[$Bug ID: Bug ID$]$, which are generated based on the clusters. For a cluster with $n$ bugs, $\frac{n*(n-1)}{2}$ duplicated pairs can be derived. Furthermore, for two clusters (with sizes n and m, respectively), we can generate $n*m$ non-duplicated pairs.

Notably, the train set is carefully balanced in terms of positive (duplicate) and negative (non-duplicate) data, while the development and test sets are intentionally left unbalanced. This setting aims to reflect the real-world scenario, where the number of non-duplicate bug pairs far exceeds the number of duplicate bugs.

\begin{table*}[h]

\caption{STATISTICS OF FIVE OPEN SOURCE DATASETS}
\fontsize{7}{9}\selectfont
\label{table:distribution of train/test set}
\label{Data Statistics}
\centering
\begin{tabular}{cccccccc}
\hline
Dataset     & Bugs           & Dup Pairs & Separate Bugs & Dup Bug Ratio & Cluster Numbers & Cluster size   \\ \hline
Eclipse     & 84020(85156)   & 13231     & 70752         & 0.1564        & 7519            & 2.760        \\
Firefox     & 96258(115814)  & 15742     & 80000         & 0.1689        & 6654            & 3.366         \\
Mozilla     & 195248(205069) & 34507     & 160378        & 0.1786        & 17263           & 2.998        \\
JDT         & 44154(45296)   & 6513      & 37608         & 0.1483        & 3828            & 2.701         \\
TBird & 24767(32551)   & 4404      & 20050         & 0.1905        & 2133            & 3.065       \\ \hline
\end{tabular}
\end{table*}
\subsection{Model Fine-tuning}
As shown in Figure \ref{Pipeline}, Model Fine-tuning phase encompasses the process of training two models: the retrieval model and the classification model.

To adapt SBERT for the retrieval task, we modify the dataset structure into triplets [Anchor, Positive, Negative], where we aim to fine-tune the model's ability to distinguish between relevant (positive) and irrelevant (negative) instances. The loss function employed is the Triplet Loss \cite{Nils2019sbert}, represented by \ref{TripletLoss}. In this equation, $||\cdot||$ denotes a distance metric used to assess the similarity between embeddings. It is important to note that this loss function imposes a condition that the distance between the anchor text and the positive text should be at least $\theta$ greater than the distance between the anchor text and the negative text.

\begin{equation}
\label{TripletLoss}
    Triplet \ Loss = Max(||E_a-E_p|| - ||E_a-E_n|| + \epsilon, 0) 
\end{equation}

In the context of classification, BERT operates by taking in two texts simultaneously, and using [SEP] token to differentiate them. The embedding of the [CLS] token, obtained from the final layer of BERT, is processed through a linear layer. The softmax function is then applied to generate the final prediction. The loss function employed in this case is the CrossEntropy (CE) Loss, as represented by \ref{CE}.

\begin{equation}
\label{CE}
    CE(y, p) = - \sum^N_{i=1}{y_i\log{p_i}}
\end{equation}

where the $y_i$ is the true label and $p_i$ is the predicted label.

\section{Experimental Settings} \label{sec:experiment}
{In this section, we detail the experiments, discussing setup, datasets, hyperparameters, evaluation, baselines, and model selection.}
\subsection{Setup}
To ensure a fair and consistent comparison between the models, we maintain uniformity by implementing and building the models using Python within the PyTorch framework. We will provide the structured code in subsequent documentation. All experimental procedures were conducted on a Linux server featuring an AMD EPYC-Rome processor and an NVIDIA A40 GPU card. This setting allows for efficient execution and reliable performance evaluation of the models.

\subsection{Datasets}
We use five open source bug report repositories \footnote{Datasets available at: [Online]. Available: \url{https://github.com/logpai/bugrepo}} to verify the effectiveness of our system, namely Eclipse, Firefox, Mozilla, JDT, and ThunderBird (TBird), as the experimental datasets in our study. These repositories have been extensively utilized in previous research \cite{gupta2021systematic}. We focus on the following statistical attributes to characterize the datasets as Table \ref{Data Statistics} shown.

We have detailed the process of Data Split in Section \ref{subsec: datasplit}, where we employ an 8:1:1 ratio to split train, development, and test sets respectively. As previously discussed, we introduce skew to the development and test data, while maintaining balance in the train set. Consequently, we adhere to the 'Dup Bug Ratio' as indicated in Table \ref{Data Statistics}, to establish the ratio of duplicate pairs in both the development and test sets. Since a substantial number of duplicate and non-duplicate pairs can be generated, we limit the size of the train/test/dev set as shown in Table \ref{table:distribution of train/test set}.

\begin{table}[h]
\centering
\caption{DISTRIBUTION OF BUG PAIRS IN TRAIN/DEV/TEST SET}
{\fontsize{7}{9}\selectfont
\label{table:distribution of train/test set}
\begin{tabular}{cccccccc}
\hline
\multirow{2}{*}{Dataset} & \multicolumn{2}{c}{Train} & \multicolumn{2}{c}{Dev} & \multicolumn{2}{c}{Test} & \multirow{2}{*}{Total} \\ \cline{2-7}
                         & Dup         & Nondup      & Dup       & Nondup      & Dup       & Nondup       &                        \\ \hline
Eclipse                  & 6615        & 6615        & 258       & 1394        & 258       & 1394         & 16534                  \\
Firefox                  & 7871        & 7871        & 332       & 1635        & 332       & 1635         & 19676                  \\
Mozilla                  & 17253       & 17253       & 770       & 3542        & 770       & 3542         & 43130                  \\
JDT                      & 3256        & 3256        & 120       & 693         & 120       & 693          & 8138                   \\
TBird              & 2202        & 2202        & 104       & 445         & 104       & 445          & 5502                   \\ \hline
\end{tabular}}
\end{table}

\begin{table*}[h]
\centering
\caption{RECALL@K OF MODELS IN DUPLICATE BUG RETRIEVAL FOR ALL DATASETS}
\label{table: single retrieval}
\fontsize{7}{9}\selectfont
\setlength{\tabcolsep}{4pt} 
\renewcommand{\arraystretch}{1.5} 
\begin{tabular}{cccccccccccccccccccccc}
\hline
         & \multicolumn{3}{c}{Eclipse}                      &           & \multicolumn{3}{c}{Firefox}                      &           & \multicolumn{3}{c}{Mozilla}                      &           & \multicolumn{3}{c}{JDT}                          &           & \multicolumn{3}{c}{TBird}                    &  &                 \\ \cline{2-4} \cline{6-8} \cline{10-12} \cline{14-16} \cline{18-20} \cline{22-22} 
         & r@20           & r@60           & r@100          &           & r@20           & r@60           & r@100          &           & r@20           & r@60           & r@100          &           & r@20           & r@60           & r@100          &           & r@20           & r@60           & r@100      &  & Avg r@100       \\ \cline{2-4} \cline{6-8} \cline{10-12} \cline{14-16} \cline{18-20} \cline{22-22} 

Fasttext & 0.489          & 0.678          & 0.783          &           & 0.596          & 0.716          & 0.809          &           & 0.414          & 0.526          & 0.588          &           & 0.608          & 0.785          & 0.972          &           & 0.627          & 0.874          & \textbf{1.000} &  & 0.8304          \\
Glove    & 0.602          & 0.727          & 0.824          &           & 0.705          & 0.789          & 0.843          &           & 0.478          & 0.608          & 0.662          &           & 0.579          & 0.798          & 0.975          &           & 0.689          & 0.888          & \textbf{1.000} &  & 0.8608          \\
SBERT & \textbf{0.848} & \textbf{0.935} & \textbf{0.960} & \textbf{} & \textbf{0.892} & \textbf{0.956} & \textbf{0.973} & \textbf{} & \textbf{0.771} & \textbf{0.892} & \textbf{0.919} & \textbf{} & \textbf{0.872} & \textbf{0.990} & \textbf{0.997} & \textbf{} & \textbf{0.880} & \textbf{0.983} & \textbf{1.000} &  & \textbf{0.9698} \\ \hline
\end{tabular}
\end{table*}

\begin{table*}
\centering
\caption{PRECISION, RECALL \& F1 SCORES OF MODELS IN DUPLICATE BUG CLASSIFICATION TASK FOR ALL DATASETS}
\label{table: single classification}
\fontsize{7}{9}\selectfont
\setlength{\tabcolsep}{3.5pt} 
\renewcommand{\arraystretch}{1.5} 
\begin{tabular}{cccccccccccccccccccccc}
\hline
& \multicolumn{3}{c}{Eclipse} & & \multicolumn{3}{c}{Firefox} & & \multicolumn{3}{c}{Mozilla} & & \multicolumn{3}{c}{JDT} & & \multicolumn{3}{c}{TBird} & & \\
\cline{2-4} \cline{6-8} \cline{10-12} \cline{14-16} \cline{18-20} \cline{22-22}
& Precision & Recall & F1 & & Precision & Recall & F1 & & Precision & Recall & F1 & & Precision & Recall & F1 & & Precision & Recall & F1 & & Avg F1 \\
\cline{2-4} \cline{6-8} \cline{10-12} \cline{14-16} \cline{18-20} \cline{22-22}
Bi-LSTM & 0.511 & 0.506 & 0.473 & & 0.510 & 0.515 & 0.469 & & 0.507 & 0.506 & 0.506 & & 0.490 & 0.490 & 0.490 & & 0.621 & 0.510 & 0.474 & & 0.4824 \\

DC-CNN & 0.752 & 0.813 & 0.785 & & 0.744& 0.765 & 0.753 & & 0.792 & 0.765 & 0.736 & & 0.763 & 0.781 & 0.773 & & 0.833 & 0.752 & 0.781 & & 0.7660 \\
BERT  & \textbf{0.825} & 0.888 & 0.848 & & 0.881 & 0.921 & 0.899 & & 0.824 & \textbf{0.892} & 0.849 & & 0.772 & 0.857 & 0.797 & & 0.870 & 0.898 & 0.883 & & 0.8552 \\
ALBERT & 0.806 & \textbf{0.896} & 0.834 & & 0.874 & 0.920 & 0.893 & & 0.819 & 0.889 & 0.845 & & \textbf{0.825} & \textbf{0.872} & \textbf{0.843} & & \textbf{0.885} & \textbf{0.902} & \textbf{0.893} & & 0.8616 \\
RoBERTa & 0.846 & 0.892 & 0.866 & & \textbf{0.886} & \textbf{0.925} & \textbf{0.903} & & \textbf{0.835} & 0.891 & \textbf{0.857} & & 0.824 & 0.868 & 0.841 & & 0.846 & 0.898 & 0.866 & & \textbf{0.8666} \\
\hline
\end{tabular}
\end{table*}

\subsection{Hyperparameters\label{sec:hypara}} 
We leverage pre-trained transformer-based models along with their respective tokenizers. Fine-tuning of these models is performed using the AdamW optimizer \cite{Ilya2017Adam} with a learning rate of $10^{-5}$. 

In the classification scenario with SBERT, we introduce a linear layer comprising two hidden layers of 768 hidden size each. 
For the Bi-LSTM model, we utilize the SGD optimizer, implementing a learning rate of 0.5 and a decay rate of 0.25. 
For the CNN model, we adhere to the configurations outlined in DC-CNN \cite{he2020duplicate}.

To mitigate overfitting, we apply a 0.5 dropout across all models. We process training data in 32-size batches. To bolster the robustness and reliability of the results, each experiment is conducted five times.

\subsection{Evaluation}

\subsubsection{Individual Evaluation}
The performance of retrieval and classification models is individually assessed in our study. For the retrieval model, we evaluate the performance by measuring the recall and precision under different Top-k settings. For the classification model, we employ precision, recall, and the corresponding F1 score to indicate the performance.

\textbf{Metrics}: Utilizing a confusion matrix, which tabulates the counts of True Positives (TP), False Positives (FP), False Negatives (FN), and True Negatives (TN), we characterize the recall, precision, and F1 score as delineated in \ref{recall}, \ref{precision}, and \ref{f1_score} respectively. Above fomulas are the performance indicators for classification. However, in the context of retrieval, the computation of recall@k and precision@k deviates slightly, as demonstrated in \ref{recall@k} and \ref{precision@k}.

\begin{equation}
\label{recall}
    Recall = TP / (TP + FN)
\end{equation}

\begin{equation}
\label{precision}
    Precision = TP / (TP + FP)
\end{equation}

\begin{equation}
\label{f1_score}
    F1 = \frac{2 * (Precision * Recall)}{(Precision + Recall)}
\end{equation}

\begin{equation}
    \label{recall@k}
    Recall@k = \frac{(relevant \ items \ in \ top-k)}{(relevant \ items)}
\end{equation}

\begin{equation}
    \label{precision@k}
    Precision@k = \frac{(relevant \ items \ in \ top\-k)}{k}
\end{equation}

\subsubsection{Architecture Evaluation}
We conduct a comprehensive evaluation of our proposed system, comparing its performance and efficiency against individual retrieval and classification methods in two common real-world scenarios.

\textbf{One VS All}: In this scenario, when a user enters a bug, the system compares the user's input bug with existing bug reports in the entire database. To evaluate this scenario, we divided the test set into two parts: 20\% for user input and 80\% for the database. 

\textbf{All VS All}: This scenario often arises on the database side, where developers need to locate and eliminate all duplicate errors in the database. It resembles the One VS All scenario, except that we utilize the entire test set as the database.

The applications of our proposed system as well as  retrieval and classification methods in the above scenarios are as follows:

One VS All scenario: when a user submits a bug report,

\begin{itemize}
    \item \textbf{retrieval} scans the entire database to find the report most similar to the submitted report; 
     \item \textbf{classification} predicts which reports in the database are relevant to the submitted report based on certain characteristics;
     \item \textbf{proposed system} firstly retrieves the top K most similar reports from the database based on user submissions, and then the classifier further predicts these K reports to ultimately determine the duplicate results.
     
\end{itemize}
All VS All scenario: each method repeats the One VS All process, aiming to identify and eliminate all duplicates in the database.

In our approach, the classification process primarily serves to enhance the quality of retrieval results, so using the retrieved metrics allows for a more comprehensive comparison of overall performance, while also displaying the improvements attained by the classification part.
Therefore, this evaluation consists of the following metrics: recall@k, precision, accuracy (as depicted by \ref{accuracy}) and inference time. 

\begin{equation}
    \label{accuracy}
    Accuracy = TP + TN / TP + TN + FP + FN
\end{equation}
It should be noted that the maximum k set in our experiments is 100 which builds also in the real word scenario. In practical information retrieval settings, users rarely browse beyond the top 100 results due to the huge amount of data and the limitation of their own attention. Therefore, capping k at 100 strikes a balance between presenting enough relevant results and preventing users from being overwhelmed by too many results.

\begin{figure*}[t]
  \centering
  \begin{subfigure}[b]{0.49\textwidth}
    \centering
    \includegraphics[width=\textwidth]{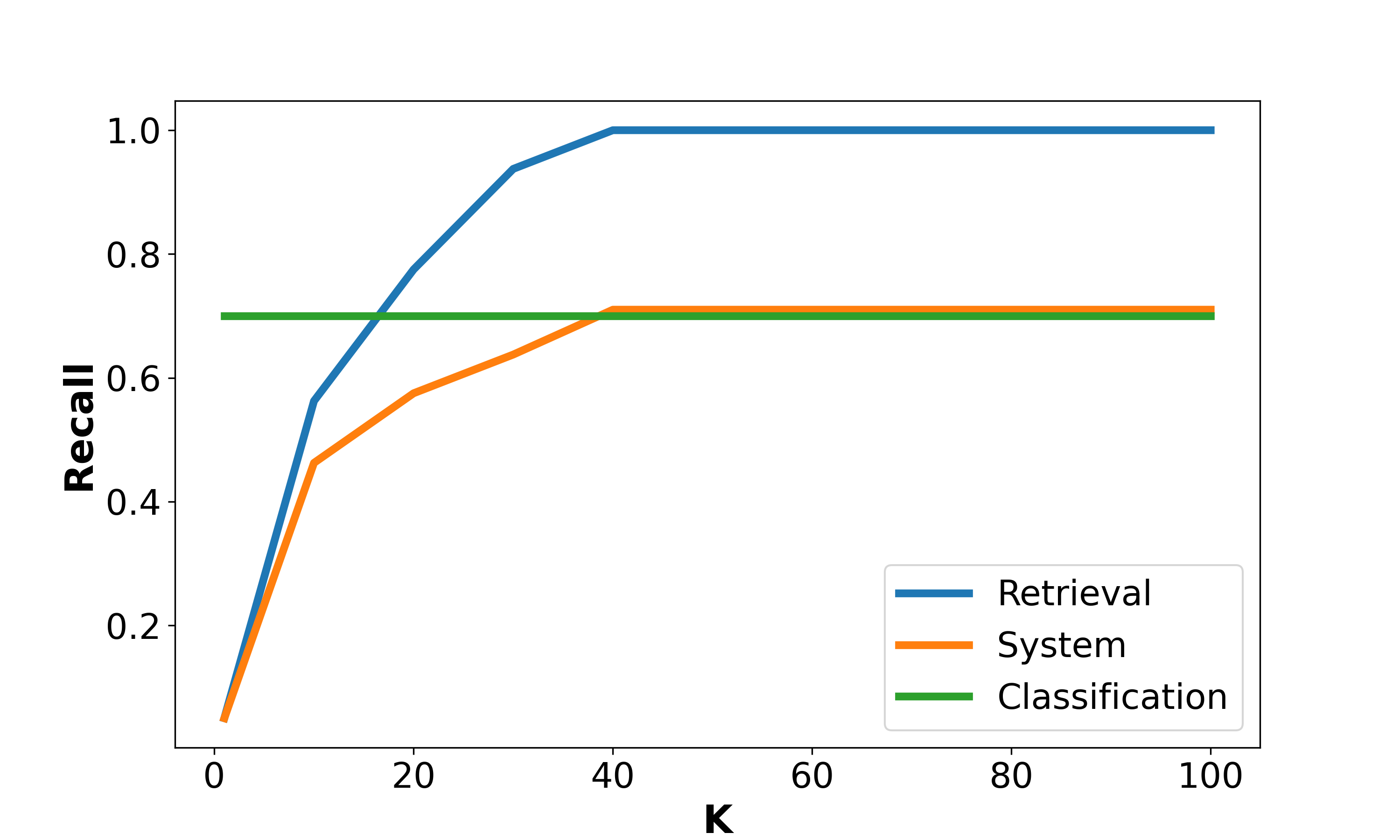}
    \caption{Recall}
    \label{subgraph: recall validation}
  \end{subfigure}
    \hfill
  \begin{subfigure}[b]{0.49\textwidth}
    \centering
    \includegraphics[width=\textwidth]{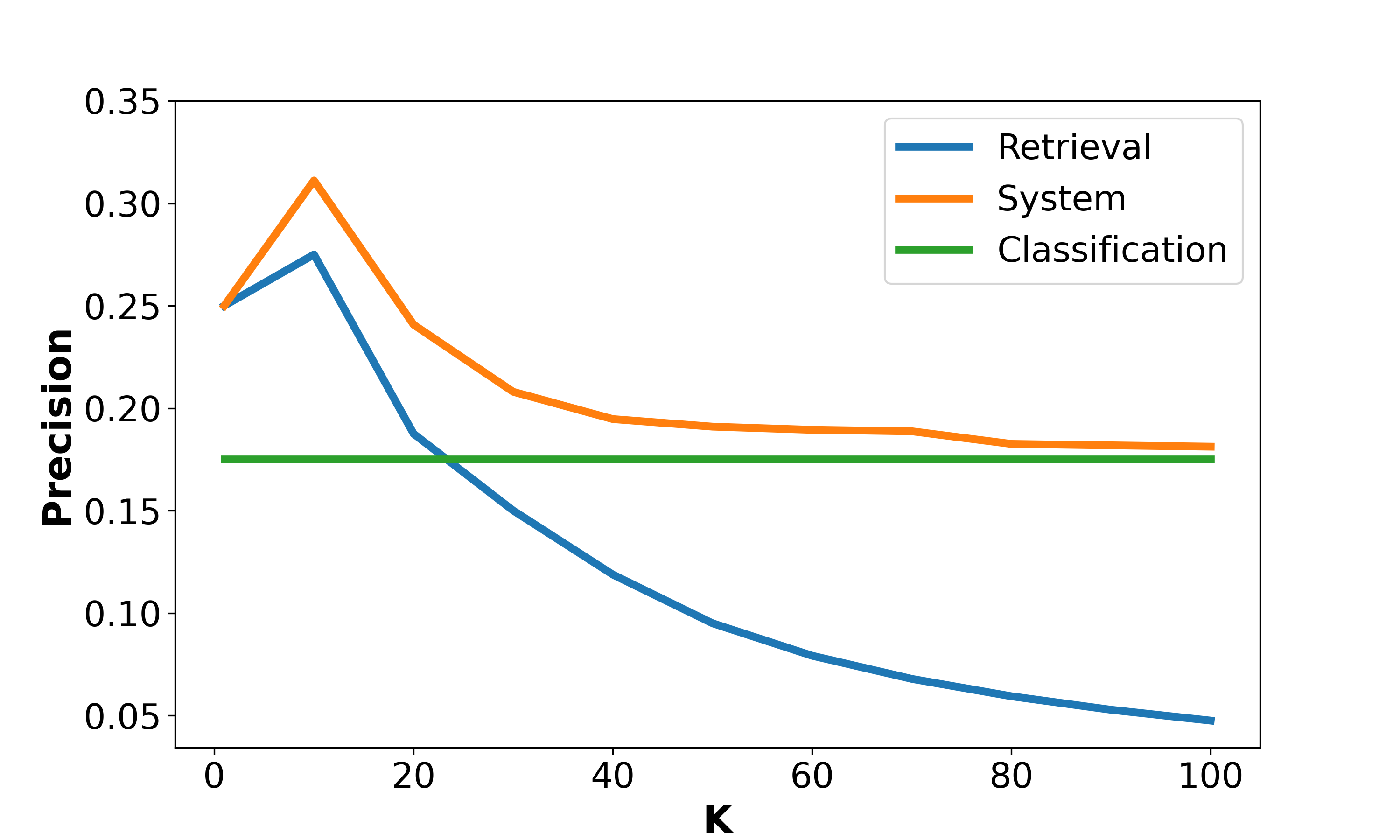}
    \caption{Precision}
    \label{subgraph: Precision validation}
    \label{fig:figure1}
  \end{subfigure}
    \hfill
  \begin{subfigure}[b]{0.49\textwidth}
    \centering
    \includegraphics[width=\textwidth]{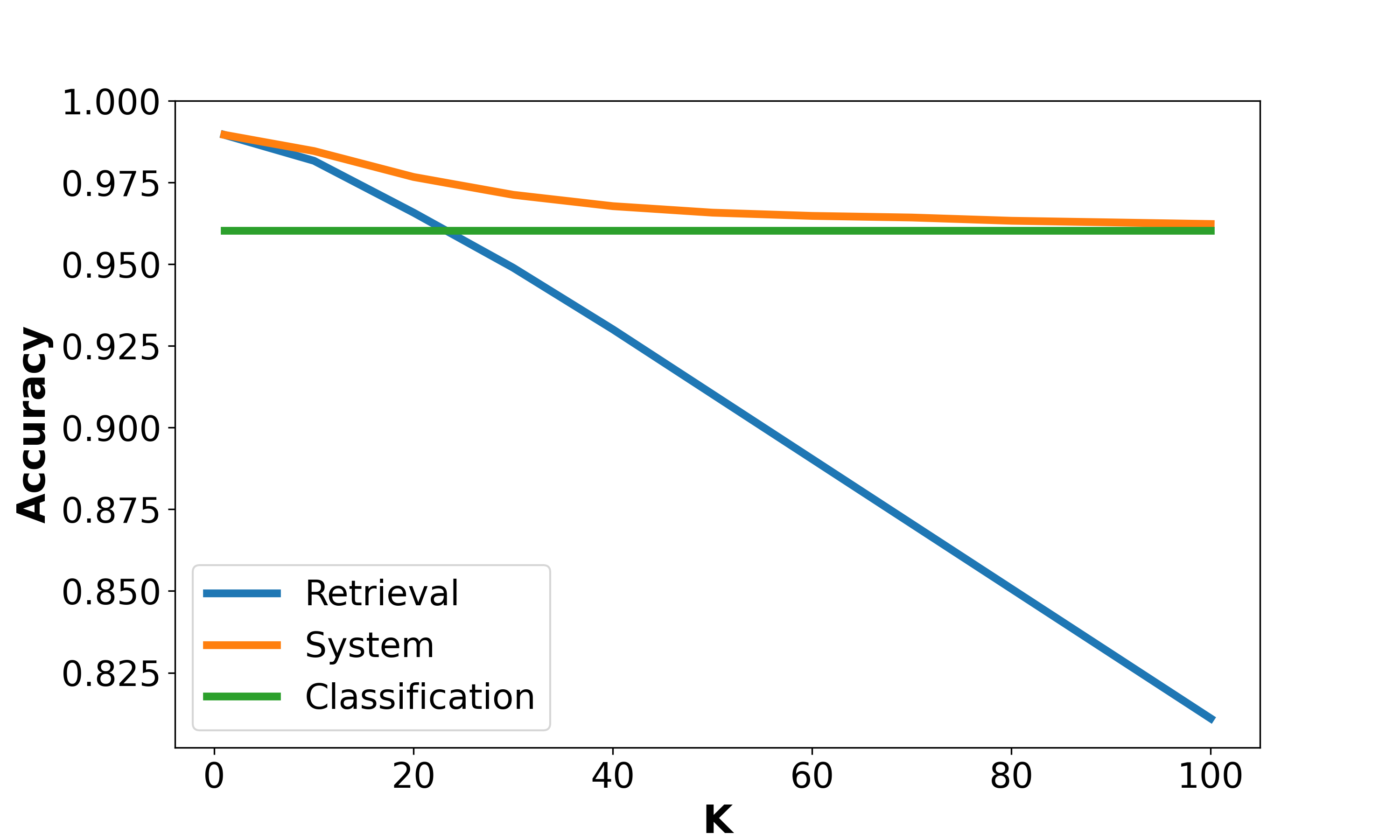}
    \caption{Accuracy}
    \label{subgraph: Accuracy validation}
  \end{subfigure}
    \hfill
  \begin{subfigure}[b]{0.49\textwidth}
    \centering
    \includegraphics[width=\textwidth]{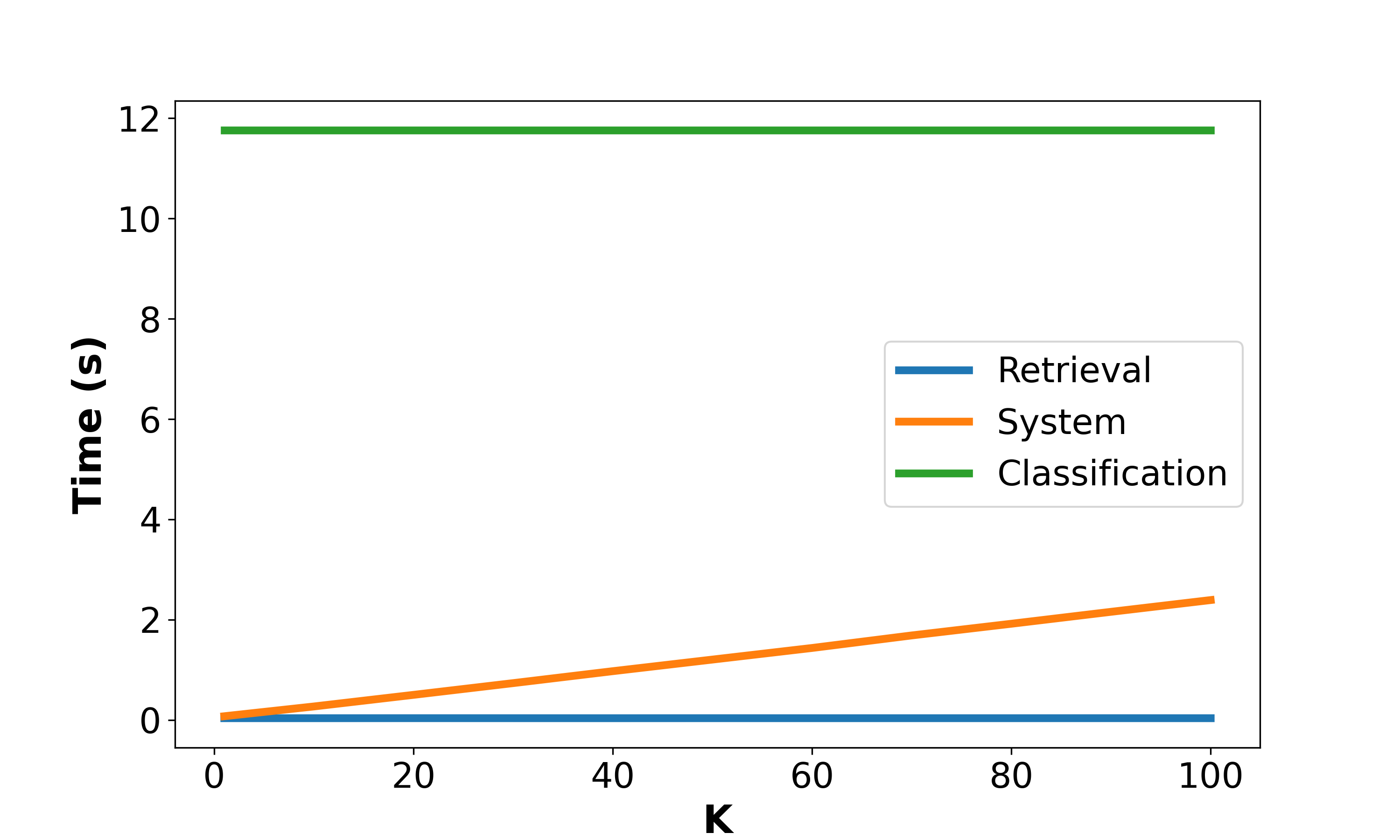}
    \caption{Running Time}
    \label{subgraph: Time validation}
  \end{subfigure}
  \caption{Time-Performance Evaluation on Firefox Dataset in One VS All scenario}
  \label{fig:Time OVA}
\end{figure*}

\subsection{Baselines}
Based on individual evaluation in retrieval and classification respectively, we employ GloVe \cite{pankajakshan2021glove} and FastText \cite{Piotr2016fasttext} as retrieval baseline methods. Since both methods generate word-level embeddings, we compute sentence embeddings by averaging the word embeddings. In the evaluation of classification models, we incorporate the Bi-LSTM and DC-CNN as the baselines. Both the Bi-LSTM model and the CNN model have been previously applied as classification models in DBRD research \cite{he2020duplicate}\cite{deshmukh2017towards}. 

In the overall evaluation, we select outperforming retrieval and classification models as baselines and compare them with our combined approach. 

\subsection{Model Selection} 
In our proposed approach, we select transformer-based models for retrieval and classification. For the retrieval model, we leverage sentence BERT (SBERT) to generate text embeddings and evaluate its efficacy on the retrieval task. In the classification task, we choose three transformer-based models as classification models, BERT, ALBERT and RoBERTa, and compare their performance. We selected these transformer-based models because they have demonstrated effectiveness in previous state-of-the-art studies \cite{Messaoud2022Bert}\cite{rocha2021siamese}\cite{isotani2021duplicate} for DBRD.

\section{Experiment Results} \label{sec:result}

We analyze our experimental results by answering following two research questions. \\

\textbf{RQ1: Compared to baseline models, how do the transformer-based models perform on retrieval and classification?}

In our first evaluation, we conduct experiments on retrieval models and classification models, presenting the results in Table \ref{table: single retrieval} and Table \ref{table: single classification}, respectively. 

Consistent with previous research, our evaluation of the retrieval model primarily focuses on the recall value. Notably, as the k value increased, we observed a substantial improvement in the recall of the model. This result is expected as increasing the value of k allows for more candidates to be considered, thereby raising the probability of identifying duplicate bugs. Upon setting k to 100, we discovered that nearly all duplicates are successfully detected, resulting in an approximate recall value of 1. 

Furthermore, by comparing the retrieval performance of the three models, as shown in Table \ref{table: single retrieval}, we find that SBERT achieves the highest recall value, under different k values. It outperforms Glove and FastText in all five datasets by an average lead of 12.42\%. This significantly demonstrates the superiority of the transformer model in information retrieval capabilities.

In the classification task, we compared the performance of traditional models, such as Bi-LSTM, DC-CNN, with transformer-based models on the three indicators of precision, recall and f1. Our results in Table \ref{table: single classification} show that f1 is significantly improved by 20\% to 38\% when using transformer-based models compared to traditional methods. When comparing transformer models, their performance does not exhibit significant variations. However, RoBERTa has emerged as the frontrunner, surpassing the others with a slightly higher F1 score of 0.8666. 

Therefore, the above experimental results indicate that transformer-based models outperform traditional models in both classification and retrieval performance.\\

\textbf{RQ2: Compared to single retrieval and classification model, how does the proposed system perform in case of recall precision, accuracy and time?}

\begin{figure*}[t]
  \centering
  \begin{subfigure}[b]{0.49\textwidth}
    \centering
    \includegraphics[width=\textwidth]{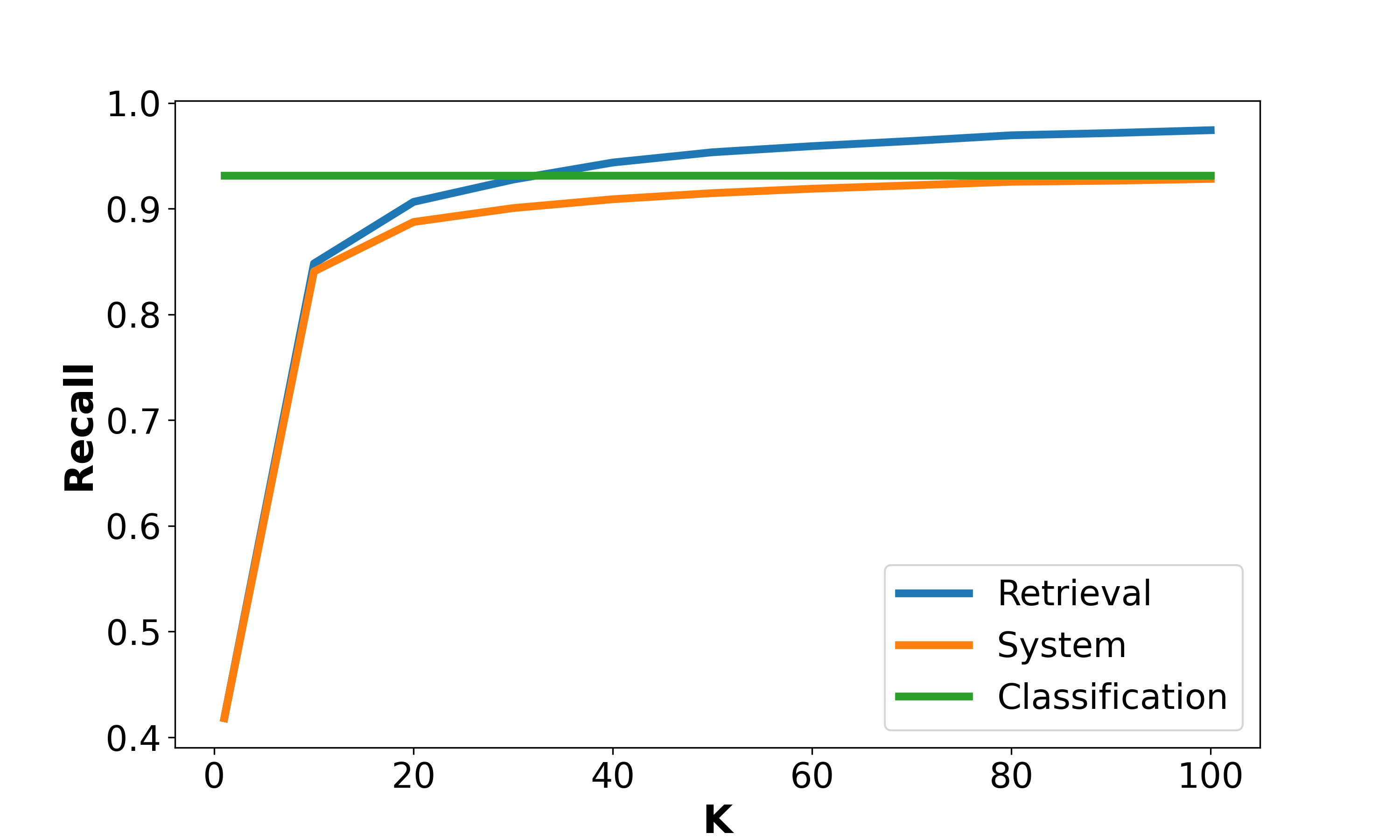}
    \caption{Recall}
    \label{subgraph1: recall validation}
  \end{subfigure}
    \hfill
  \begin{subfigure}[b]{0.49\textwidth}
    \centering
    \includegraphics[width=\textwidth]{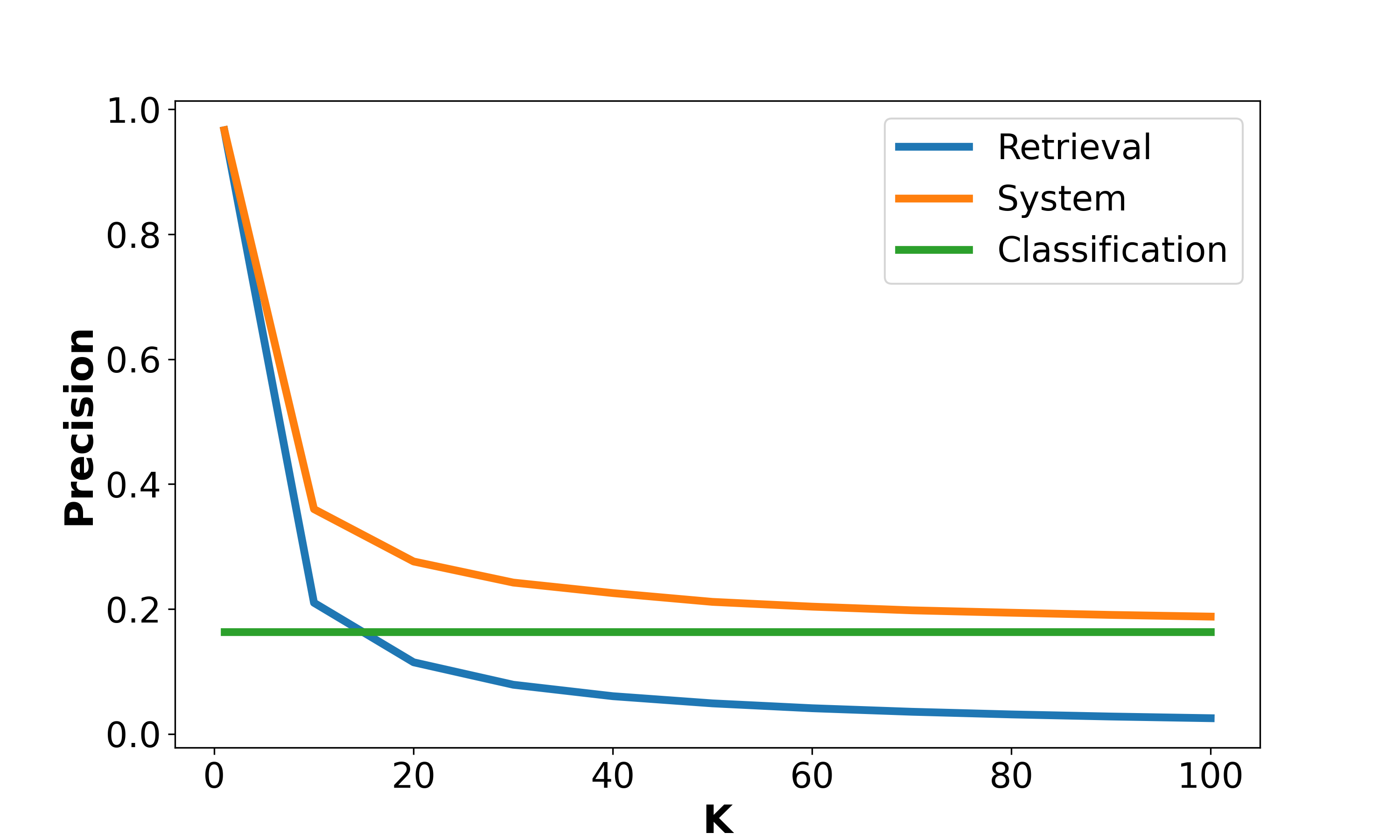}
    \caption{Precision}
    \label{subgraph1: Precision validation}
  \end{subfigure}
    \hfill
  \begin{subfigure}[b]{0.49\textwidth}
    \centering
    \includegraphics[width=\textwidth]{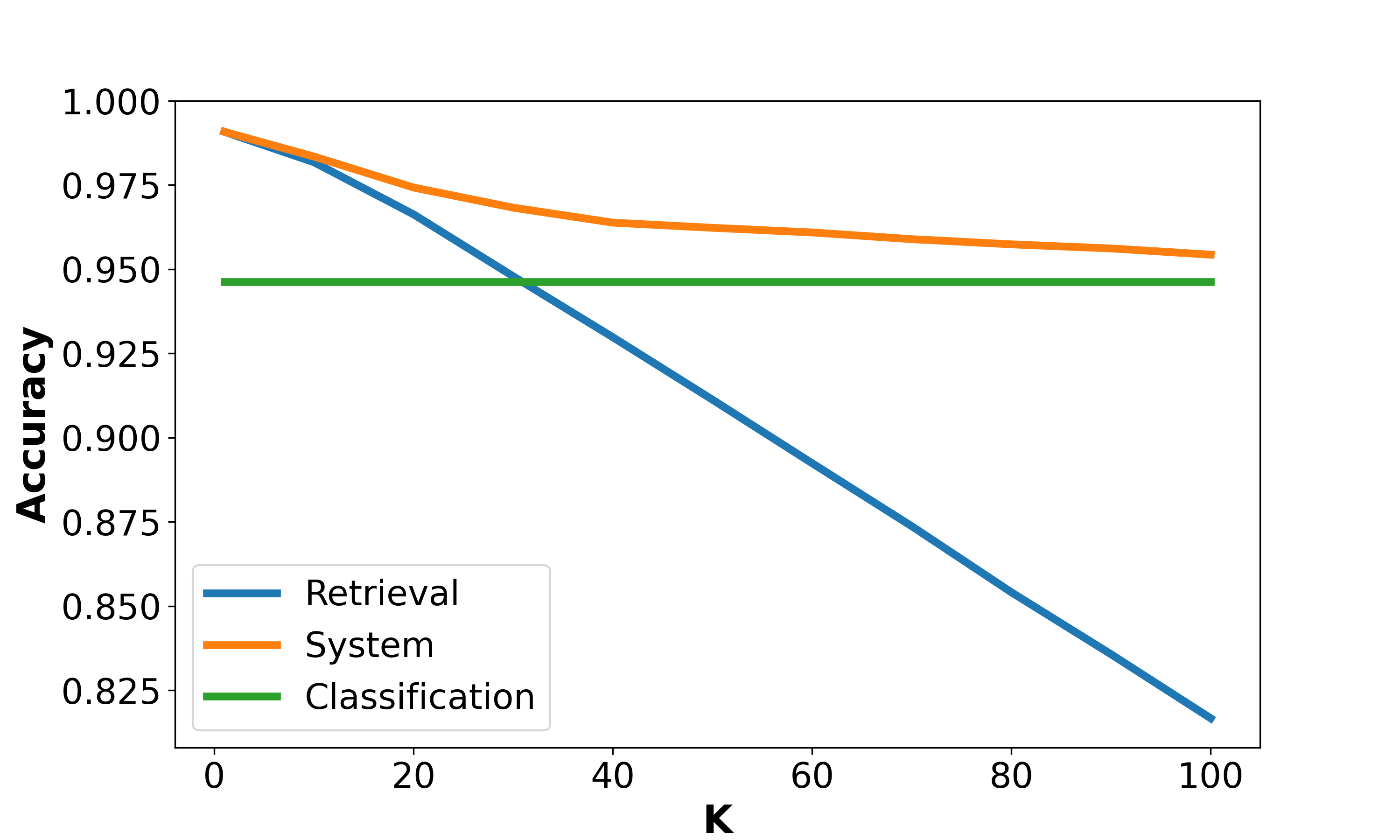}
    \caption{Accuracy}
    \label{subgraph1: Accuracy validation}
  \end{subfigure}
    \hfill
  \begin{subfigure}[b]{0.49\textwidth}
    \centering
    \includegraphics[width=\textwidth]{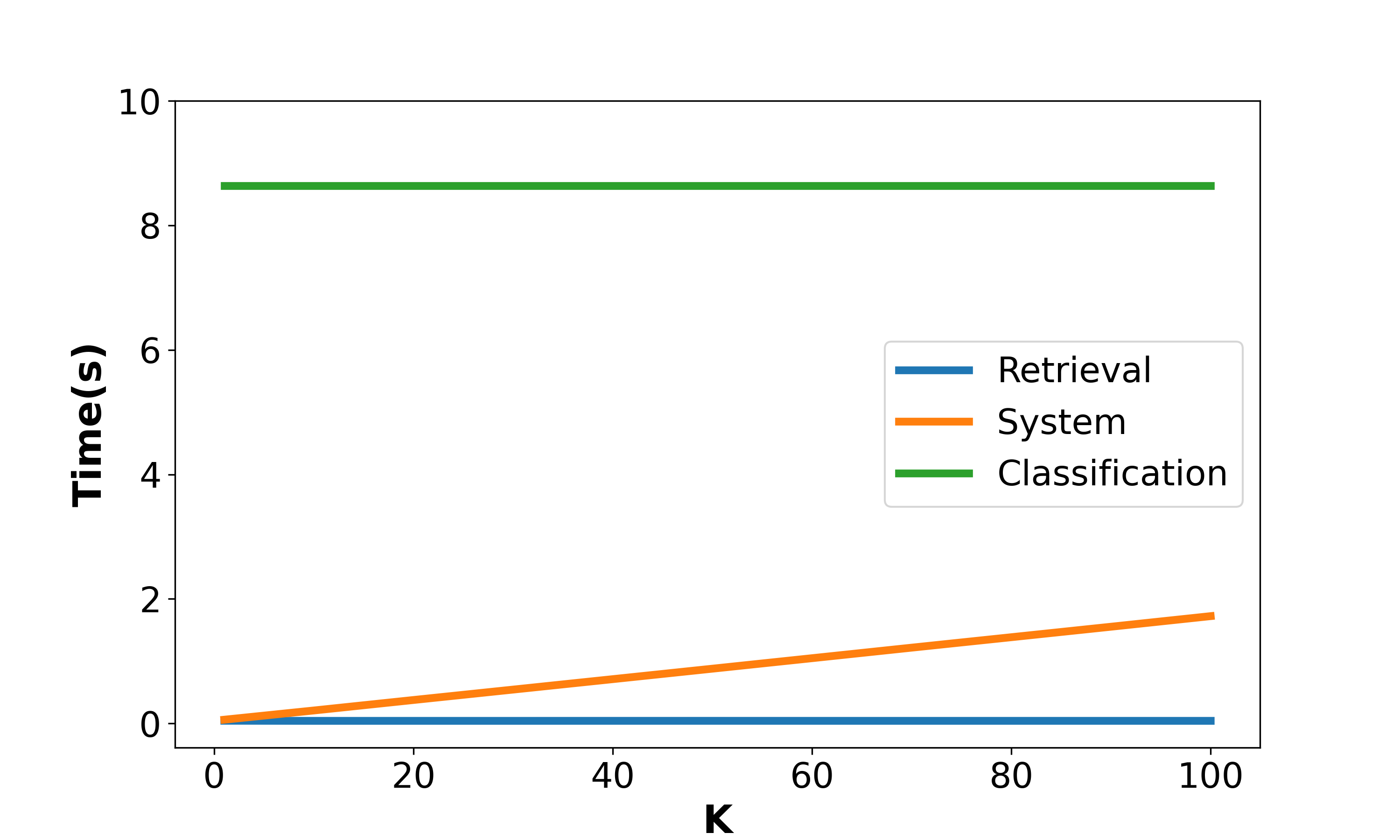}
    \caption{Running Time}
    \label{subgraph1: Time validation}
  \end{subfigure}
  \caption{Time-Performance Evaluation on Firefox Dataset in All VS All scenario}
  \label{fig:ALL}
\end{figure*}

Figure \ref{fig:Time OVA} presents a comparative performance overview showing the difference in recall, precision, accuracy and running time for single retrieval model, classification model and proposal system in the One VS All scenario, and Figure \ref{fig:ALL} presents the overall performance in the All VS All scenario. The results obtained in the One vs All and All VS All scenarios of the five datasets are relatively similar, so we choose one of datasets, firefox, to show the results. 

It is important to emphasize that, as shown in Figure \ref{fig:Time OVA}, the performance of the classification model is not affected by changes in k, thus presenting a horizontal line in the figure.

In Figure \ref{subgraph: recall validation}, we observed that at lower k, both the retrieval model and our system exhibit lower recall scores compared to the classification model. The reason is that the limited k prevents the retrieval of a large number of duplicate pairs. However, as k gradually increases, after reaching around 20, the recall of the retrieval model exceeds that of the classification model. At the same time, since the classification step of the system may introduce positive and false samples, the recall rate of the system is lower than that of the retrieval model, resulting in a decrease in the overall recall rate. Given that our system incorporates retrieval, such recall aligns with the rising trend demonstrated by retrieval models as k increases, albeit at a slightly slower pace. For example, it is not until "k" equals 40 that the recall of our method starts to be comparable to that of classification. This suggests that as the value of k rises to higher values (e.g., over 100), our recall will continue to rise, thereby establishing an increasingly discernible gap from the recall of classification.

In the Figure \ref{subgraph: Precision validation}, we note that the precision of the retrieval model and our proposed system decrease as k increases, a consequence of increasing the number of retrieved candidates. Nevertheless, it is worth mentioning that compared with the retrieval precision, the decrease of system precision is not obvious. Even when k is equal to 100, the precision of the system is still higher than the classification precision, which demonstrates the effectiveness of our system in terms of precision. Similar to precision in Figure \ref{subgraph: Accuracy validation}, accuracy also exhibits a sonsistent trend. As the k increases, both the accuracy of the system and retrieval decrease. The decline in system accuracy is also slower compared to retrieval accuracy. This highlights the advantage of our system for introducing a classification step after retrieval, as it can efficiently preserve the performance in precision and accuracy, especially better than classification.

Comparatively, as displayed in Figure \ref{subgraph: Time validation}, the classification process is more time-consuming than both retrieval and our system. This observation can be explained through a simple calculation. Assuming we have $n$ user input bugs and $m$ database bugs, and assuming that both the classification model and the retrieval model require the same time for a single inference, the following holds: The classification model requires $n * m$ inferences. The retrieval model requires $n + m$ inferences, along with $n * m$ calculations of embeddings similarity and subsequent sorting. As the similarity calculation and sorting are considerably faster than model inference, the retrieval process roughly takes n + m seconds per inference. Our system incorporates the results of retrieval. Therefore, it includes the retrieval model inference time $n + m$, similarity calculation, and sorting time, followed by $n * topk$ classifications. Consequently, the system's required time amounts to $n + m + n * topk$ inferences. The preceding calculations are equally applicable to the All VS All scenario as well. It is clear that our method exhibits almost the same remarkable efficiency as retrieval and classification in terms of time. This is proven by the fact that the time consumed by the classification is approximately 60 times greater than our approach.

Figure \ref{fig:ALL} shows that the All VS All scenario exhibits similar performance trends as the One VS All scenario. The running time in Figure \ref{subgraph1: Time validation} represents the average time taken to match each bug with its similar ones, comparable to Figure \ref{subgraph: Time validation}.


%

Overall, our system makes a trade-off by sacrificing some running time in order to maintain robust performance in terms of recall, precision, and accuracy. As k increases from 0 to 100, the recall of our system increases, demonstrating the ability of our model to successfully retrieve all relevant duplicates. At the same time, the accuracy and precision are only slightly reduced, effectively alleviating the sharp decline in retrieval, but never fall below those achieved by classification. The steady improvement in recall between retrieval and classification, coupled with the maintained superior precision, shows that we minimize the time cost while maintaining accuracy performance. This convincingly demonstrates our ability to obtain a trade-off between accuracy performance and time efficiency.

Therefore, selecting the appropriate value for k requires careful consideration. While a smaller k value may improve the time efficiency, it may also lead to a degradation in model performance. On the other hand, choosing a larger k value may result in increased time consumption. Thus, striking a balance between speed and model performance depends on selecting the optimal k value.

\section{Conclusion and Future work}
In our work, we proposed a novel system based on the transformer models, that leverages the strengths of both retrieval and classification approaches for duplicate bug report detection task. We have evaluated the transformer-based models employed by our method on five datasets, demonstrating their effectiveness compared to traditional models for both classification and retrieval. More importantly, our method shows a competitive edge by achieving a balance between time efficiency and accuracy, compared to solutions employing only one of them. This advantage holds significant importance in real-time bug report detection where requires high-quality results in a short time. In other words, under resource constraints, combining retrieval and classification as a novel solution enables dynamic adjustments to efficiently address issues related to changes in data volume and quality, while flexibly adapting to time-sensitivity and shifts in user demands. This approach enhances resource efficiency and ensures the maintenance of response speed and accuracy in a constantly changing environment. Furthermore, our combined strategy can be expanded to tackle similar issues in other tasks, such as recommendation tasks.

While our system addresses the running time concern that previous methods overlooked, and achieves a trade-off between time and accuracy, there are several factors need to be considered for practical application, such as the size of the model. Our system relies on both the retrieval and classification models, resulting in a larger memory space requirement compared to a single model. As a result, future efforts could explore the possibility of employing multi-task learning to integrate these two models, allowing for the completion of both tasks with a single model simultaneously. Additionally, there are some limitations in our study that can be addressed in the future, such as expanding to new datasets. This not only includes datasets that are more current, but also those that are more diverse in terms of types, which would enhance the generalizability of our methods.

\section*{Acknowledgment}
This work was funded by the China Scholarship Council (CSC) and supported by the Leiden Institute of Advanced Computer Science (LIACS).

\bibliographystyle{IEEEtran}
\bibliography{IEEEexample}

\end{document}